\documentclass[a4paper]{article}

\usepackage{useful}

\usepackage{tikz}

\usepackage{tikz-3dplot}
\usepackage{pgfplots}
\usepgfplotslibrary{groupplots}

\usepgfplotslibrary{external}
\usepackage{booktabs}
\usepackage{siunitx}

\usepackage{uclcolor}
\definecolor{dark-gray}{gray}{0.3}

\usepackage{subcaption}

\usepackage[backend=bibtex,citestyle=authoryear-comp,%
  bibstyle=authoryear-comp,uniquename=false,useprefix=true]{biblatex}
\usepackage{algorithm,algorithmic}

\newcommand{\haus}{\mathcal{H}}
\newcommand{\sphere}{\mathbb{S}}
\newcommand{\stiefel}{\mathbb{V}}
\newcommand{\orthonormal}{\mathbb{O}}

\newcommand{\dnorm}{\mathsf{N}}
\newcommand{\dunif}{\mathsf{U}}

\title{Geodesic Monte Carlo on Embedded Manifolds}
\author{Simon Byrne and Mark Girolami\\ \\
  Department of Statistical Science\\University College London}

\begin{document}
\maketitle
\begin{abstract}
  \noindent
  Markov chain Monte Carlo methods explicitly defined on the manifold of
  probability distributions have recently been established. These methods are
  constructed from diffusions across the manifold and the solution of the
  equations describing geodesic flows in the Hamilton--Jacobi
  representation. This paper takes the differential geometric basis of Markov
  chain Monte Carlo further by considering methods to simulate from
  probability distributions that themselves are defined on a manifold, with
  common examples being classes of distributions describing directional
  statistics. Proposal mechanisms are developed based on the geodesic flows
  over the manifolds of support for the distributions and illustrative
  examples are provided for the hypersphere and Stiefel manifold of orthonormal
  matrices.

  \smallskip
  \noindent \textbf{Keywords:} directional statistics, geodesic, Hamiltonian
  Monte Caro, Riemannian manifold, Stiefel manifold.
\end{abstract}

\section{Introduction}
Markov chain Monte Carlo (MCMC) methods which originated in the physics
literature have caused a revolution in statistical methodology over the last
20 years by providing the means, now in an almost routine manner, to perform
Bayesian inference over arbitrary non-conjugate prior and posterior pairs of
distributions \parencite{gilks1996}.

A specific class of MCMC methods, originally known as Hybrid Monte Carlo
(HMC), was developed to more efficiently simulate quantum chromodynamic
systems \parencite{duane1987}. HMC goes beyond the random walk Metropolis or
Gibbs sampling schemes and overcomes many of their shortcomings. In particular
HMC methods are capable of proposing bold long distance moves in the state
space which will retain a very high acceptance probability and thus improve
the rate of convergence to the invariant measure of the chain and reduce the
autocorrelation of samples drawn from the stationary distribution of the
chain.  The HMC proposal mechanism is based on simulating Hamiltonian dynamics
defined by the target distribution, see \textcite{neal2011} for a
comprehensive tutorial. For this reason HMC is now routinely referred to as
Hamiltonian Monte Carlo. Despite the relative strengths and attractive
properties of HMC it has largely been bypassed in the literature devoted to
MCMC and Bayesian statistical methodology with very few serious applications
of the methodology being published.

More recently \textcite{girolami2011} defined a Hamiltonian scheme that is
able to incorporate geometric structure in the form a Riemannian metric.
The Riemannian manifold Hamiltonian Monte Carlo (RMHMC) methodology makes
proposals implicitly via Hamiltonian dynamics on the manifold defined by the
Fisher--Rao metric tensor and the corresponding Levi--Civita connection. The
paper has raised an awareness of the differential geometric foundations of
MCMC schemes such as HMC and has already seen a number of methodological and
algorithmic developments as well as some impressive and challenging
applications exploiting these geometric MCMC
methods \parencite{martin2012,vanlier2012,raue2012,ender2011}.

In contrast to \textcite{girolami2011}, in this particular paper we show how
Hamiltonian Monte Carlo methods may be designed for and applied to
distributions defined on manifolds embedded in Euclidean space, by exploiting
the existence of explicit forms for geodesics. This can provide a significant
boost in speed, by avoiding the need to solve large linear systems as well as
complications arising due to the lack of a single global coordinate system.

By way of specific illustration we consider two such manifolds: the unit
hypersphere, corresponding to the set of unit vectors in $\R^d$, and its
extension to Stiefel manifolds, the set of $p$-tuples of orthogonal unit
vectors in $\R^d$. Such manifolds occur in many statistical applications:
distributions on circles and spheres, such as the von Mises distribution, are
common in problems dealing with directional
data \parencite{mardia2000}. Orthonormal bases arise in dimension-reduction
methods such as factor analysis \parencite{jolliffe1986}, and can be used to
construct distributions on matrices via eigendecompositions.

The problem of sampling from such distributions has not received much
attention. Most methods in wide use, such as those used in directional
statistics for sampling from spheres, have been developed for the specific
problem at hand, often based on rejection sampling techniques tuned to a
specific family. For the various multivariate extensions of these
distributions, these techniques are usually embedded in a Gibbs sampling
scheme.

There are relatively few works on the general problem of sampling from
manifolds. The recent paper by \textcite{diaconis2012} provides a readable
introduction to the concepts of geometric measure theory, and practical issues
when sampling from manifolds, with the motivation of computing certain
sampling distributions for hypothesis testing.  \textcite{brubaker2012},
somewhat similar to our approach, develop a HMC algorithm using the
iterative algorithm for approximating the Hamiltonian paths.

In the next section, we provide a brief overview of the necessary
concepts from differential geometry and geometric measure theory, such as
geodesics and Hausdorff measures. In section~\ref{sec:hmc-manif} we construct
a Hamiltonian integrator that utilises the explicit form of the geodesics, and
incorporate this into a general HMC
algorithm. Section~\ref{sec:embedded-manifolds} gives examples of various
manifolds for which the geodesic equations are known, and
section~\ref{sec:examples} provides some illustrative applications.

\section{Manifolds, geodesics and measures}
\label{sec:defs}

\subsection{Manifolds and embeddings}
\label{sec:manifolds}

In this section, we introduce the necessary terminology from differential
geometry and information geometry. A more rigorous treatment can be found in
reference books such as \textcite{docarmo1976,docarmo1992} and \textcite{amari2000}.

An $m$-dimensional manifold $\mathcal{M}$ is a set that
locally acts like $\R^m$: that is, for each point $x \in \mathcal{M}$, there
is a bijective mapping $q$, called a \emph{coordinate system}, from an open
set around $x$ to an open set in $\R^m$. Our particular focus is on manifolds
that are \emph{embedded} in some higher-dimensional Euclidean space $\R^n$,
(that is, they are submanifolds of $\R^n$). Note that $\R^d$ is itself a
$d$-dimensional manifold, which we refer to as the \emph{Euclidean manifold}.

\begin{example}
  \label{exm:sphere}
  A simple example of an embedded manifold is the \emph{hypersphere} or
  \emph{$(d-1)$-sphere}:
  \begin{equation*}
    \sphere^{d-1} = \bigl\{ x \in \R^d : \|x\| = 1 \bigr\}.
  \end{equation*}
  This is a $(d-1)$-dimensional manifold, as there exists an angular coordinate
  system $\phi \in (0,2\pi) \times (0,\pi)^{d-2}$ where
  \begin{align*}
    x_1 &= \sin \phi_1 \ldots \sin \phi_{n-2} \sin \phi_{n-1}, \\
    x_2 &= \sin \phi_1 \ldots \sin \phi_{n-2} \cos \phi_{n-1}, \\
    x_3 &= \sin \phi_1 \ldots \cos \phi_{n-2}, \\
    & \ \vdots \\
    x_{n-1} &= \sin \phi_1 \cos \phi_2, \\
    x_{n} &= \cos \phi_1 .
  \end{align*}
  Note that this coordinate system excludes some points of $\sphere^{d-1}$:
  such as $\delta_d = (0,\ldots,0,1)$.  As a result, it is not a \emph{global}
  coordinate system (in fact, no global coordinate system for $\sphere^{d-1}$
  exists),  nevertheless it is possible to cover all of $\sphere^{d-1}$ by
  utilising  multiple coordinate systems known as an \emph{atlas}.
\end{example}

A \emph{tangent} at a point $x \in \mathcal{M}$ is a vector $v$ that lies ``flat'' on the manifold.
More precisely, it can be defined as an equivalence class of the set of functions
$\{\gamma:[a,b] \to \mathcal{M} : \gamma(t_0) = x\}$ that have the same ``time
derivative'' $\D{t} q(\gamma(t))|_{t=t_0}$ in some coordinate system $q$. 
For an embedded manifold, however, a tangent can be represented simply as a vector $v \in \R^n$ such that
\begin{equation*}
  v = \dot\gamma(t_0) = \D{t} \gamma(t) \bigr|_{t=t_0}.
\end{equation*}
The \emph{tangent space} is the set $T_x$ of such vectors, and form a subspace
of $\R^n$: this is equal to the span of the set of partial derivatives
$\Dp[x_i]{q_j}$ of some coordinate system $q$.
\begin{example}
  \label{exm:sphere-tangent}
  A function on the sphere $\gamma:[a,b] \to \sphere^{d-1}$, must satisfy the
  constraint $\sum_i [\gamma_i(t)]^2 = 1$. By taking the time derivative of
  both sides, we find that
  \begin{equation*}
    \D{t} \sum_{i=1}^d [\gamma_i(t)]^2 = 2 \sum_{i=1}^d \gamma_i(t)
    \dot\gamma_i(t) = 0.
  \end{equation*}
  Therefore the tangent space at $x \in \sphere^{d-1}$ is the
  $(d-1)$-dimensional subspace of vectors orthogonal to $x$:
  \begin{equation*}
    T_x = \{v \in \R^d : x^\top v = 0\}.
  \end{equation*}
\end{example}

A \emph{Riemannian manifold} incorporates a notion of distance, such that for a
point $q \in \mathcal{M}$, there exists a positive-definite matrix $G$, called
the metric tensor, that forms an inner product between tangents $u$ and $v$
\begin{equation*}
  \langle u, v \rangle_G = u^\top G(q) v.
\end{equation*}

\emph{Information geometry} is the application of differential geometry to
families of probability distributions. Such a family $\{p(\cdot \mid \theta)
: \theta \in \Theta\}$ can be viewed as a Riemannian manifold,
using the \emph{Fisher--Rao metric tensor}
\begin{equation*}
  G_{ij} = - E_{X \mid \theta} \left[ \DDtp{\theta_i}{\theta_j} \log p(X \mid
    \theta) \right] .
\end{equation*}

\begin{example}
  \label{exm:fisher-multinomial}
  The family of $d$-dimensional multinomial distributions
  \begin{equation*}
    p(z \mid \theta) = \theta_1^{z_1} \cdot \ldots \cdot \theta_d^{z_d} ,
    \quad z = \delta_1, \ldots, \delta_d
  \end{equation*}
  where $\delta_i$ is the $i$th coordinate vector, is parametrised by the unit
  $(d-1)$-simplex,
  \begin{equation*}
    \Delta^{d-1} = 
    \{ \theta \in \R^d : \theta_i \geq 0, \sum_j \theta_j = 1\}.
  \end{equation*}
  This is a $(d-1)$-dimensional manifold embedded in $\R^n$, and can be
  parametrised in $(d-1)$ dimensions by dropping the last element of $\theta$,
  the set of which we will denote by $\Delta^{d-1}_{(-d)}$.

  The Fisher--Rao metric tensor in $\Delta^{d-1}_{(-d)}$ is then easily shown
  to be
  \begin{equation*}
    G_{ij} 
    \begin{cases}
      \frac{1}{\theta_i} + \frac{1}{1-\sum_{k=1}^{d-1} \theta_k} 
        & \quad \text{if } i=j,\\
      \frac{1}{1-\sum_{k=1}^{d-1} \theta_k} &\quad \text{otherwise}.
    \end{cases}
  \end{equation*}  
\end{example}

A smooth mapping from a Riemannian manifold to $\R^n$ is an \emph{isometric
  embedding} if the Riemannian inner product is equivalent to the usual
Euclidean inner product. That is
\begin{equation*}
  s^\top G(q) t = u^\top v, 
  \qquad \text{where } 
  u_i = \sum_j \Dp[x_i]{q_j} s_i, \ v_i = \sum_j \Dp[x_i]{q_j} t_i,
\end{equation*}
or equivalently,
\begin{equation}
  \label{eq:iso-embed}
  G_{ij} = \sum_{l=1}^d \Dp[x_l]{q_i} \Dp[x_l]{q_j}.
\end{equation}
The existence of such embeddings is determined by the celebrated
\textcite{nash1956} embedding theorem, however it doesn't give any guide as
how to construct them. Nevertheless, there are some such embeddings we can identify.

\begin{figure}
  \centering
  \tdplotsetmaincoords{70}{110}
  \begin{tikzpicture}[tdplot_main_coords,scale=2.5]

    \draw[->] (0,0,0) -- (1.2,0,0);
    \draw[->] (0,0,0) -- (0,1.2,0);
    \draw[->] (0,0,0) -- (0,0,1.2);

    \draw[fill=midgreen!50] (1,0,0) -- (0,1,0) -- (0,0,1) -- cycle;

    \foreach \m in {0.1,0.2,...,0.9} {
      \draw (1-\m,\m,0) -- (1-\m,0,\m);
      \draw (0,1-\m,\m) -- (\m,1-\m,0);
      \draw (\m,0,1-\m) -- (0,\m,1-\m);
    }
  \end{tikzpicture}
  \quad
  \tdplotsetmaincoords{70}{110}
  \begin{tikzpicture}[tdplot_main_coords,scale=2.5]

    \draw[->] (0,0,0) -- (1.2,0,0);
    \draw[->] (0,0,0) -- (0,1.2,0);
    \draw[->] (0,0,0) -- (0,0,1.2);

    \tdplotsetthetaplanecoords{0}
    \begin{scope}
      \draw[clip] (1,0,0) arc (0:90:1) 
      {\pgfextra{\tdplotsetthetaplanecoords{90}} 
        [tdplot_rotated_coords] arc (90:0:1)} 
      {[tdplot_rotated_coords]  arc (0:90:1)};
      ;
      \shade[ball color=white] (0,0,0) circle [radius=1cm];
    \end{scope}

    \foreach \m in {0.1,0.2,...,0.9}  {
      \pgfmathsetmacro{\sqrtm}{sqrt(\m)}
      \pgfmathsetmacro{\sqrtom}{sqrt(1-\m)}
      \draw (0,\sqrtm,\sqrtom) arc (90:0:\sqrtm);
      \tdplotsetthetaplanecoords{0}
      \draw[tdplot_rotated_coords] (0,\sqrtm,\sqrtom) arc (90:0:\sqrtm);
      \tdplotsetthetaplanecoords{90}
      \draw[tdplot_rotated_coords] (0,\sqrtm,-\sqrtom) arc (90:0:\sqrtm);
    }
  \end{tikzpicture}
  \caption{Unit 2-simplex $\Delta^2$ and the positive orthant of the 2-sphere
    $\sphere^2$. The lines on the simplex are equidistant: the transformation
    to the sphere stretches these apart near the boundary.}
  \label{fig:simplex-sphere}
\end{figure}
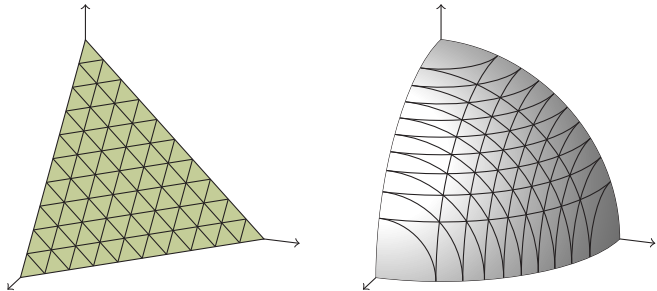

\begin{example}
  \label{exm:simplex-sphere-embedding}
  There is a bijective mapping from the simplex $\Delta^{d-1}$  to the
  positive orthant of the sphere $\sphere^{d-1}$ by taking the element-wise
  square root $x_i = \sqrt{\theta_i}$ (see
  Figure~\ref{fig:simplex-sphere}). If we consider it as a mapping from
  $\Delta^{d-1}_{(-d)}$, the partial derivatives are of the form
  \begin{equation*}
    \Dp[x_l]{\theta_i} =
    \begin{cases}
      \tfrac{1}{2} \theta_i^{-1/2} & \quad \text{if } i = l < d, \\ 
      0 &\quad \text{if } i \neq l < d, \\
      \tfrac{1}{2} (1-\sum_{k=1}^{d-1} \theta_k)^{-1/2} & \quad \text{if }
      i=l=d .
    \end{cases}
  \end{equation*}
  Note that by \eqref{eq:iso-embed}, this is an isometric embedding (up to
  proportionality) of the Fisher--Rao metric from Example~\ref{exm:fisher-multinomial}.
\end{example}

\subsection{Geodesics}

The \emph{affine connection} of a manifold determines the relationship between
tangent spaces of different points on a manifold: interestingly, this depends
on the path $\gamma:[a,b] \to \mathcal{M}$ used to connect the two points, and
for a vector field $v(t) \in T_{\gamma(t)}$ along the path, we can measure the
change by the \emph{covariant derivative}. 

Of course, the time derivative $\dot\gamma(t) = \D[\gamma(t)]{t}$ is itself a
such a vector field: when this follows the affine connection, the covariant
derivative is 0, in which case $\gamma$ is known as a \emph{geodesic}.

This property can be expressed by the geodesic equation
\begin{equation}
  \label{eq:geodesic-eqn}
  \ddot\gamma_i(t) + \sum_{j,k} \Gamma_{jk}^i\bigl(\gamma(t) \bigr) \dot\gamma_j(t) \dot\gamma_k(t) = 0 ,
\end{equation}
where $\Gamma_{jk}^i(x)$ are known as the connection coefficients or Christoffel
symbols. A Riemannian manifold induces a natural affine connection known as
the \emph{Levi-Civita connection}.

In the Euclidean manifold $\R^n$, the Christoffel symbols $\Gamma_{jk}^i $ are zero, and so the
geodesic equation \eqref{eq:geodesic-eqn} reduces to $\ddot\gamma(t) =
0$. Hence the geodesics are the set of straight lines $\gamma(t) = a t + b$.

In a Riemannian manifold, the geodesics are
the locally extremal paths (maxima or minima in terms of calculus of
variations) of the integrated path length
\begin{equation*}
  \int_a^b \|\dot\gamma(t)\|_G \, \d t ,
  \quad \text{where }
  \|v\|^2_G = v^\top G v.
\end{equation*}
Moreover, the geodesics have \emph{constant speed}, in that
$\|\dot\gamma(t)\|_G$ is constant over $t$. As the geodesics can be
determined by the metric, they are consequently preserved under any metric-preserving
transformation, such as an isometric embedding.

\begin{figure}
  \centering
  \tdplotsetmaincoords{70}{110}
  \begin{tikzpicture}[tdplot_main_coords,scale=2.3]
    
    \shade[ball color=white] (0,0,0) circle [radius=1cm];
    
    \draw (0,0,0) circle [radius=1];
    
    \foreach \phiv in {15,30,...,75} {
      \pgfmathsetmacro\sinphiv{sin(\phiv)}%
      \pgfmathsetmacro\cosphiv{cos(\phiv)}%
      \draw[dark-gray] (0,0,\sinphiv) circle [radius=\cosphiv];
      \draw[dark-gray] (0,0,-\sinphiv) circle [radius=\cosphiv];
    }
    
    \foreach \thetav in {0,15,...,165} {
      \tdplotsetthetaplanecoords{\thetav}
      \draw[dark-gray,tdplot_rotated_coords] (0,0,0) circle [radius=1];
    };

    \tdplotsetrotatedcoords{60}{-60}{0} 
    
    \draw[tdplot_rotated_coords,darkred,thick,*->] 
    (-90:1) 
    arc (-90:0:1) node[circle,inner sep=1pt] {};
    
    \draw[tdplot_rotated_coords,darkred,dashed] 
    (-90:1) arc (-90:270:1) ;

    \tdplotsetrotatedcoords{60}{90}{0} 
    \draw[tdplot_rotated_coords,thick] 
    (0,0,1) node[circle,fill=black,inner sep=1pt]{}
    circle [radius=0.2];

    \tdplotsetrotatedcoords{60}{60}{0} 
    \draw[tdplot_rotated_coords,thick] 
    (0,0,1) node[circle,fill=black,inner sep=1pt]{}
    circle [radius=0.2];

    \tdplotsetrotatedcoords{60}{30}{0} 
    \draw[tdplot_rotated_coords,thick] 
    (0,0,1) node[circle,fill=black,inner sep=1pt]{}
    circle [radius=0.2];

  \end{tikzpicture}
  \quad
  \begin{tikzpicture}
    \begin{axis}[scale=0.65,
      xlabel=$\phi_2$,ylabel=$\phi_1$,
      y label style={rotate=-90},
      xtick={-180,0,180},xticklabels={$0$,$\pi$,$2\pi$},
      minor xtick={-165,-150,...,165},
      ytick={-90,90},yticklabels={$0$,$\pi$},
      minor ytick={-75,-60,...,75},
      grid=both,
      ymin=-90,ymax=90,xmin=-180,xmax=180,
      ]

      \addplot[domain=-180:180,darkred,samples=100,thick,dashed] 
      {asin(sqrt(3)*sin(x)/sqrt(1+3*sin(x)^2))}; 
      \addplot[domain=-1:90,darkred,samples=100,thick,*->] 
      {asin(sqrt(3)*sin(x)/sqrt(1+3*sin(x)^2))}; 

      \draw[thick] (axis cs:90,0) 
      node[circle,fill=black,inner sep=1pt]{}
      circle [x radius=11.5,y radius=11.5];

      \draw[thick] (axis cs:90,30) 
      node[circle,fill=black,inner sep=1pt]{}
      circle [x radius=13.3, y radius=11.5];

      \draw[thick] (axis cs:90,60)
      node[circle,fill=black,inner sep=1pt]{}
      circle [x radius=23, y radius=11.5];

    \end{axis}
  \end{tikzpicture}
  \caption{A geodesic (\linesample{darkred,*->}) and great circle (\linesample{darkred,dashed}) on the sphere $\sphere^2$, and its path in
    the spherical polar coordinate system
    $x = (\sin\phi_1 \sin\phi_2, \sin\phi_1 \cos\phi_2, \cos\phi_1)$. The
    ellipses correspond to equi-length tangents from each marked point.}
  \label{fig:sphere-geodesic}
\end{figure}
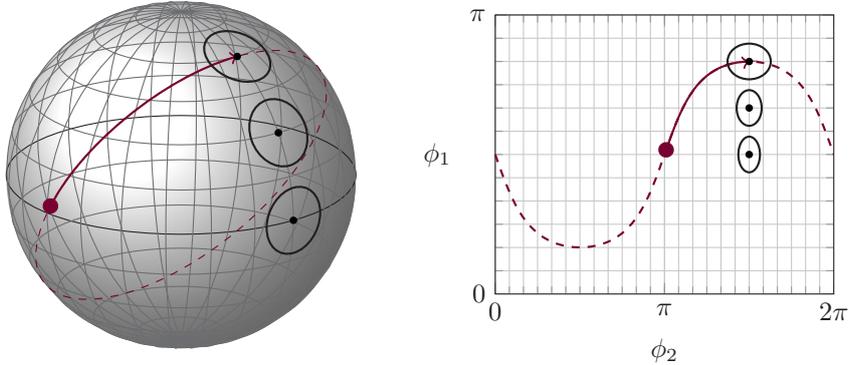

\begin{example}
  \label{exm:sphere-geodesic}
  A standard result in differential geometry is that the geodesics of the
  $n$-sphere are rotations about the origin, known as \emph{great circles}
  (see Figure~\ref{fig:sphere-geodesic}):
  \begin{equation*}
    x(t) = x(0) \cos (\alpha t) + \frac{v(0)}{\alpha} \sin (\alpha t)
  \end{equation*}
  where $x(0) \in \sphere^n$ is the initial position, $v(0)$ is the initial
  velocity in the tangent space (\ie such that $x(0)^\top v(0) = 0$), and
  $\alpha = \|v(0)\|$ is the constant angular velocity.
\end{example}

For any geodesic $\gamma:[a,b] \to \mathcal{M}$, the \emph{geodesic flow} describes the path of the
geodesic and its tangent $\bigl(\gamma(t),\dot\gamma(t)\bigr)$. Moreover, it
is unique to the initial conditions $(x,v) = (\gamma(a),\dot\gamma(a))$, so
we can describe any geodesic flow from its starting position $x$ and
velocity $v$: this is also known as the \emph{exponential map}. If all such
pairs $(x,v)$ describe geodesics, the manifold is said to be
\emph{geodesically complete}, which is true of the manifolds we consider in
this paper.

\subsection{The Hausdorff measure and distributions on manifolds}

As our motivation is to sample from distributions defined on manifolds, we
introduce some basic concepts of geometric measure theory that will be useful
for this purpose. Geometric measure theory is a large and active topic, and is
covered in detail in references such as \textcite{federer1969} and
\textcite{morgan2009}. However for a more accessible overview with a
statistical flavour, we suggest the recent introduction given by
\textcite{diaconis2012}.

Our key requirement is a reference measure from which we can specify
probability density functions, similar to the role played by the Lebesgue
measure for distributions on Euclidean space. For this we use the
\emph{Hausdorff measure}, one of the fundamental concepts in
geometric measure theory. This can be defined rigorously in terms of a limit
of coverings of the manifold (see the above references), however for a
manifold embedded in $\R^n$, it can be heuristically interpreted as the surface area of the manifold.

The relationship between $\haus^m$, the $m$-dimensional Hausdorff measure, and
$\lambda^m$, the Lebesgue measure on $\R^m$, is given by the \emph{area
  formula} \parencite[Theorem 3.2.5]{federer1969}. If we parametrise the
manifold by a Libschitz function $f:\R^m \to \R^n$, then for any
$\haus^m$-measurable function $g:\R^n \to \R$,
\begin{equation*}
  \int_A g(f(u)) \, J_m f(u) \, \lambda^m (\d u) 
  = 
  \int_{R^n} g(x) \, \bigl|\{u \in A : f(u) = x\}\bigr| \, \haus^m (\d x).
\end{equation*}
Here $J_m f(x)$ is the $m$-dimensional Jacobian of $f$: this can be defined as
a norm on the matrix of partial derivatives $D f(x)$ \parencite[Section 3.2.1]{federer1969}, and if $\rank D f(x) = m$, then $[J_m f(x)]^2$ is equal
  to the sum of squares of the determinants of all $m \times m$ submatrices of $D f(x)$.

\begin{example}
  \label{exm:simplex-sphere-jacobian}
  The square root mapping in Example~\ref{exm:simplex-sphere-embedding} from
  $\Delta^{d-1}_{(-d)}$ to $\sphere^{d-1}$ has $(d-1)$-dimensional Jacobian
  \begin{equation*}
    \frac{1}{2^{d-1}} \prod_{i=1}^d \theta_i^{-1/2} .
  \end{equation*}
  The Dirichlet distribution is a distribution on the simplex, with density
  \begin{equation*}
    \frac{1}{B(\alpha)} \prod_{i=1}^d \theta_i^{\alpha_i-1}
  \end{equation*}
  with respect to the Lebesgue measure on $\Delta^{d-1}_{(-d)}$. Therefore,
  the corresponding density with respect to the Hausdorff measue on
  $\sphere^{d-1}$ is
  \begin{equation*}
    \frac{2^{d-1}}{B(\alpha)} \prod_{i=1}^d x_i^{2\alpha_i-1}.
  \end{equation*}
  In other words, the uniform distribution on the sphere arises when $\alpha_i
  = \frac{1}{2}$, whereas $\alpha=1$ gives the uniform distribution on the
  simplex (see Figure~\ref{fig:beta-circle}).
\end{example}
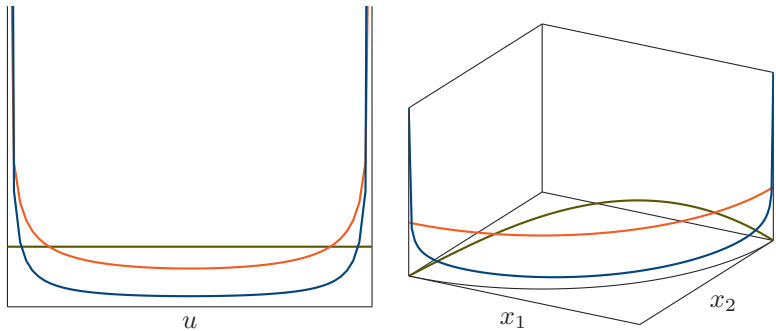
\begin{figure}
  \centering
  \begin{tikzpicture}
    \begin{axis}[scale=0.7,
      axis x line*=bottom,
      xtick=\empty,xlabel=$u$,ytick=\empty,
      ymin=0,ymax=5,xmin=0,xmax=1]
      
      \addplot[domain=0:1,darkgreen,thick] {1}; 
      \addplot[domain=0.001:0.999,midred,samples=60,thick] 
        {x^(-0.5)*(1-x)^(-0.5)/3.141593};          
      \addplot[domain=0.001:0.999,navyblue,samples=60,thick] 
        {x^(-0.9)*(1-x)^(-0.9)/19.71464};        
    \end{axis}
  \end{tikzpicture}
  \quad
  \begin{tikzpicture}
    \begin{axis}[view={30}{30},scale=0.7,
      xtick=\empty,ytick=\empty,ztick=\empty,
      xlabel={$x_1$},ylabel={$x_2$},
      zmin=0,zmax=1,xmin=0,xmax=1,ymin=-1,ymax=0]

      \addplot3[domain=0:0.5*pi,samples=60,samples y=0]
      ({sin(deg(x))},{-cos(deg(x))}, {0});

      \addplot3[domain=0.0001:0.5*pi,samples=120,samples y=0,darkgreen,thick]
      ({sin(deg(x))},{-cos(deg(x))},{abs(sin(deg(x))*cos(deg(x)))});

      \addplot3[domain=0:0.5*pi,samples=60,samples y=0,midred,thick]
      ({sin(deg(x))},{-cos(deg(x))},{1/3.141593});

      \addplot3[domain=0.0001:0.5*pi,samples=120,samples y=0,navyblue,thick]
      ({sin(deg(x))},{-cos(deg(x))},
      {min(sin(deg(x))^(-0.4)*cos(deg(x))^(-0.4)/19.71464,1)});
    \end{axis}
  \end{tikzpicture}    
  \caption{Densities of different beta$(\alpha,\alpha)$ distributions for $u
    \in (0,1)$ (left), and their corresponding transformations to the positive
    quadrant of the unit circle
    $\sphere^1$, by the mapping $u \mapsto (\sqrt{1-u},\sqrt{u})$
    (right). $\alpha = 0.1$ (\linesample{navyblue}), $\alpha = 0.5$ (\linesample{midred}), and $\alpha = 1.0$ 
    (\linesample{darkgreen}).
  }
  \label{fig:beta-circle}
\end{figure}

The area formula allows the Hausdorff measure to be easily extended to
Riemannian manifolds \parencite[Section 3.2.46]{federer1969}, where
\begin{equation*}
  \haus^m ( \d q) = \sqrt{ |G(q)|} \lambda^m(\d q) .
\end{equation*}
This construction would be familiar to Bayesian statisticians as the \emph{Jeffreys
prior}, in the case where $G$ is the Fisher--Rao metric.

When working with probability distributions on manifolds, the Hausdorff
measure forms the natural reference measure, and allows for reparametrisation
without needing to compute any additional Jacobian term. We use $\pi_\haus$ to
denote the density with respect to the Hausdorff measure of the distribution
of interest.

\begin{example}
  \label{exm:von-mises}
  The von Mises distribution is a common family of distributions defined on
  the unit circle \parencite[section 3.5.4]{mardia2000}. When parametrised by an angle $\theta$, the density with
  respect to the Lebesgue measure on $[0,2\pi)$ is
  \begin{equation*}
    \pi(\theta) = \frac{1}{2 \pi I_0(\kappa)} \exp\{\kappa \cos (\theta - \mu)\}.
  \end{equation*}
  The embedding transformation $x = (\sin \theta, \cos \theta)$ has unit
  Jacobian, so the density with respect to the $1$-dimensional
  Hausdorff measure is
  \begin{equation*}
    \pi_\haus(x) = \frac{1}{2 \pi I_0(\|c\|)} \exp\{c^\top x\},
  \end{equation*}
  where $c = (\kappa \sin \mu, \kappa\cos \mu)$ and $I_k$ is the modified
  Bessel function of the first kind. In other words, it is a natural
  exponential family on the circle.

  The von Mises--Fisher distribution is the natural extension to higher-order
  spheres \parencite[section 9.3.2]{mardia2000}, with density
  \begin{equation*}
    \pi_\haus(x) = \frac{\|c\|^{p/2 -1}}{(2 \pi)^{p/2} I_{p/2-1}(\|c\|)} \exp \{c^\top x\}.
  \end{equation*}
  Attempting to write this as a density with respect to the Lebesgue measure
  on some parametrisation of the surface, such as angular coordinates, would
  be much more involved, as the Jacobian is no longer constant.
\end{example}

\section{Hamiltonian Monte Carlo on embedded manifolds}
\label{sec:hmc-manif}

Riemannian manifold Hamiltonian Monte Carlo (RMHMC) is a Markov chain Monte
Carlo (MCMC) scheme where new samples are proposed by approximately solving a
system of differential equations describing the paths of Hamiltonian dynamics
on the manifold \parencite{girolami2011}.

The key requirement for Hamiltonian Monte Carlo is the \emph{symplectic
  integrator}. This is a discretisation that approximates the Hamiltonian
flows, yet maintains certain desirable properties of the exact solution,
namely time-reversibility and volume preservation which are necessary to
maintain the detailed balance conditions. The standard approach is to
use a \emph{leapfrog scheme}, which alternately updates the position and
momentum via first order Euler updates \parencite{neal2011}.

Given a target density $\pi(q)$ (with respect to the Lebesgue measure) in some
coordinate system $q$, Riemannian manifold Hamiltonian Monte Carlo,
\textcite{girolami2011} utilises a Hamiltonian of the form
\begin{equation*}
  H(q,p) = - \log \pi(q) + \frac{1}{2} \log |G(q)| + \frac{1}{2} p^\top
  G(q)^{-1} p ,
\end{equation*}
where $G$ is the metric tensor. This is the negative log of the joint density (respect to the Lebesgue
measure) for $(q,p)$, where the conditional distribution for
the auxiliary momentum variable $p$ is $\dnorm(0,G(q))$.

The first two terms can be combined into the negative log of the target
density with respect to the Hausdorff measure of the manifold
\begin{equation}
  \label{eq:rmh-haus}
  H(q,p) = - \log \pi_\haus(q) + \frac{1}{2} p^\top G(q)^{-1} p .
\end{equation}
By Hamilton's equations, the dynamics are determined by the system of
differential equations
\begin{align}
  \D[q]{t} &= \Dp[H]{p} = G(q)^{-1} p , \label{eq:pos-update}
  \\
  \D[p]{t} &= -\Dp[H]{q} = \nabla_q \bigl[ \log \pi_\haus(q) 
  - \frac{1}{2} p^\top G(q)^{-1} p \bigr] . \label{eq:mom-update}
\end{align}
As this Hamiltonian is not separable (that is, it cannot be written as the sum
of a function of $q$ and a function of $p$), we are unable to
apply the standard leapfrog integrator. 

\subsection{Geodesic integrator}
\label{sec:integrator}

\textcite{girolami2011} develop a generalised leapfrog scheme, which involves
composing adjoint Euler approximations to \eqref{eq:pos-update} and
\eqref{eq:mom-update} in a reversible manner. Unfortunately, some of these
steps do not have an explicit form, and so need to be solved implicitly by
fixed-point iterations. Furthermore, these updates require computation of both the inverse
and derivatives of the metric tensor, which are $O(m^3)$ operations; this
limits the feasibility of numerically naive implementations of this scheme for higher-dimensional
problems. Finally, such a scheme assumes a global coordinate system, which may
cause problems for manifolds for which none exist, such as the sphere, where
artificial boundaries may be induced.

In this contribution we instead construct an integrator by
\emph{splitting the Hamiltonian} \parencite[section II.5]{hairer2006}: that is, we
treat each term in \eqref{eq:rmh-haus} as a distinct Hamiltonian, and
alternate simulating between the exact solutions. 

Splitting methods have been used in other contexts to develop alternative
integrators for Hamiltonian Monte Carlo \parencite[section 5.5.1]{neal2011},
such as extending HMC to infinite-dimensional Hilbert
spaces \parencite{beskos2011}, and defining schemes that may reduce computational cost \parencite{shahbaba2012}.

We take the first component of the splitting to be the ``potential'' term
\begin{equation*}
  H^{[1]}(q,p) = - \log \pi_\haus(q).
\end{equation*}
Hamilton's equations give the dynamics
\begin{equation*}
  \dot q = \Dp[H^{[1]}]{p} = 0
  \quand
  \dot p = -\Dp[H^{[1]}]{q} = \nabla_q \log_\haus \pi(q).
\end{equation*}
Starting at $\bigl(q(0),p(0)\bigr)$, this has the exact solution
\begin{equation}
  \label{eq:potential-solution}
  q(t) = q(0) 
  \qquand
  p(t) = p(0) + t \nabla_q \log \pi_\haus(q) \bigr|_{q=q(0)} .
\end{equation}
In other words, this is just a linear update to the momentum $p$. 

The second component is the ``kinetic'' term
\begin{equation}
  \label{eq:hamil-split-2}
  H^{[2]}(q,p) = \frac{1}{2} p^\top G(q)^{-1} p .
\end{equation}
This is simply a Hamiltonian absent of any potential term, the solution of
Hamilton's equations can be easily shown to be a geodesic flow under the
Levi-Civita connection of $G$ \parencite[Theorem 3.7.1]{abraham1987}, or to be
more precise, a co-geodesic flow $\bigr(q(t),p(t)\bigr)$, where $p(t) =
G(q(t)) \dot q(t)$.

Thus, if we are able to exactly compute the geodesic flow, we can construct
an integrator by alternately simulating from the dynamics
of $H^{[1]}$ and $H^{[2]}$ for some time step $\epsilon$. Each iteration of
the integrator consists of the following steps, starting at position $(q,p)$
in the phase space
\begin{enumerate}
\item Update according to the solution to $H^{[1]}$ in
  \eqref{eq:potential-solution}, for a period of $\epsilon/2$ by setting
  \begin{equation}
    \label{eq:momentum-update}
    p \leftarrow p + \frac{\epsilon}{2} \nabla_q \log \pi_\haus (q) 
  \end{equation}

\item Update according to $H^{[2]}$, by following the geodesic flow starting
  at $(q,p)$, for a period of $\epsilon$.

\item Update again according to $H^{[1]}$ for a period of $\epsilon/2$ by
  \eqref{eq:momentum-update}.

\end{enumerate}
As $H^{[1]}$ and $H^{[2]}$ are themselves Hamiltonian systems, their solutions
are necessarily both reversible and symplectic. As the integrator is
constructed by their symmetric composition, it also will be reversible and symplectic.

Therefore the overall transition kernel for our Hamiltonian Monte Carlo scheme
from an initial position $q_0$, is as follows
\begin{enumerate}
\item Propose an initial momentum $p_0$ from $N\bigl(0,G(q_0) \bigr)$.
\item Map $(q_0,p_0) \mapsto (q_T,p_T)$ by running $T$ iterations of the above integrator.
\item Accept the $q_{T}$ as the new value with probability
  \begin{equation*}
    1 \wedge \exp\bigl\{-H(q_T,p_T) + H(q_0,p_0) \bigr\} ,
  \end{equation*}
  otherwise return the original value $q_0$.
\end{enumerate}
As with the RMHMC algorithm, the metric $G$ need only be known up to
proportionality: scaling is equivalent to changing the time step $\epsilon$.

\subsection{Embedding coordinates}
The algorithm can also be written in terms of an embedding, which avoids
altogether the computation of the metric tensor and the possible lack of a
global coordinate system. 

Given an isometric embedding $\xi:\mathcal{M} \to \R^n$, then the path $x(t) = \xi(q(t))$,
such that
\begin{equation*}
  \dot x_i (t) = \sum_j \Dp[x_i]{q_j} \dot q_j (t)
\end{equation*}
Therefore, we can transform the phase space $(q,p)$, where $\dot q = G^{-1}
p$, to the embedded phase space $(x,v)$, such that
\begin{equation*}
  v = \dot x = M G(q)^{-1} p = M (M^\top M)^{-1} p
  \quad \text{where }
  M_{ij} = \Dp[x_i]{q_j},
\end{equation*}
since $G = M^\top M$, from \eqref{eq:iso-embed}.

By substitution, the Hamiltonian \eqref{eq:rmh-haus} can be written in terms of these coordinates as
\begin{equation}
  \label{eq:hamiltonian-xv}
  H = - \log \pi_\haus(x) + \tfrac{1}{2} v^\top v.
\end{equation}
Note the target density $\pi_\haus$ is still defined with respect to
the Hausdorff measure of the manifold, and so no additional log-Jacobian term
is introduced.

We can rewrite the solution to $H^{[1]}$ in
\eqref{eq:potential-solution} in these coordinates. 
The position $x(t)$ remains constant, and by the change of variables the operator
$\nabla_q = M^\top \nabla_x$, the velocity has a linear path
\begin{equation*}
  v(t) = v(0) + t M (M^\top M)^{-1} M^\top \nabla_x \log \pi_\haus (x)
  \bigr|_{x= x(0)} .
\end{equation*}
The linear operator $M (M^\top M)^{-1} M^\top$ is the ``hat matrix'' from
linear regression: this is the orthogonal projection onto the span of the columns
of $M$, \ie the tangent space of the embedded manifold.

Although it is possible to compute this projection using standard least squares
algorithms, it can be computationally expensive and prone to numerical
instability at the boundaries of
the coordinate system (for example, at the poles of a sphere). However for all
the manifolds we consider, there exists an explicit form for an orthonormal
basis $N$ of the \emph{normal} to the tangent space, in which case we can simply
subtract the projection onto the normal:
\begin{equation*}
  v(t) = v(0) + t (I - N N^\top) \nabla_x \log \pi_\haus (x)
  \bigr|_{x= x(0)} .
\end{equation*}

Finally, we require a method for sampling the initial
velocity $v_0$. Since $p_0 \sim \dnorm(0,G(q))$, it follows that
\begin{equation*}
  v_0 \sim \dnorm(0, M (M^\top M)^{-1} M^\top) = \dnorm(0, I - N N^\top).
\end{equation*}
We don't need to compute a Cholesky decomposition here: since $(I - N N^\top)$
is a projection, it is idempotent, so we can draw $z$ from $\dnorm(0,I_n)$, and
project $v_0 = (I-N N^\top) z$ to obtain the necessary sample.

\begin{algorithm}
\begin{algorithmic}[1]
    \STATE $v \sim \dnorm(0,I_n)$
    \STATE $v \leftarrow v - N(x) N(x)^\top v$
    \STATE $h \leftarrow \log\pi_\haus(x) - \tfrac{1}{2} v^\top v$
    \STATE $x^* \leftarrow x$
    \FOR{$\tau = 1,\ldots,T$}
      \STATE $v \leftarrow v + 
        \frac{\epsilon}{2} \nabla_{x^*} \log \pi_\haus (x^*)$ 
      \STATE $v \leftarrow v - N(x) N(x)^\top v$
      \STATE Update $(x^*,v)$ by following the geodesic flow for a time interval
        of $\epsilon$
      \STATE $v \leftarrow v + 
        \frac{\epsilon}{2} \nabla_{x^*} \log \pi_\haus (x^*)$ 
      \STATE $v \leftarrow v - N(x) N(x)^\top v$  
    \ENDFOR
    \STATE $h^* \leftarrow \log\pi_\haus(x^*) - \tfrac{1}{2} v^\top v$
    \STATE $u \sim \dunif(0,1)$
    \IF{$u < \exp(h^* - h)$} 
      \STATE $x \leftarrow x^*$
    \ENDIF
\end{algorithmic}
\caption{The transition kernel for Hamiltonian Monte Carlo on an embedded manifold using geodesic flows.}
\label{alg:geohmc}
\end{algorithm}

The resulting procedure is presented in Algorithm~\ref{alg:geohmc}. In order
to implement it for an embedded manifold $\mathcal{M} \subseteq \R^n$, we need
to be able to evaluate the following at each $x \in \mathcal{M}$:
\begin{itemize}
\item The log-density with respect to the Hausdorff measure $\log \pi_\haus$,
  and its gradients.
\item An orthogonal projection from $\R^n$ to the tangent space of
  $x \in \mathcal{M}$
\item The geodesic flow from any $v \in T_x \mathcal{M}$.
\end{itemize}

Note that by working entirely in the embedded space, we completely avoid the
coordinate system $q$, and the related problems where no single global
coordinate system exists. The Riemannian metric $G$ only appears in the Jacobian
determinant term of the density: in certain examples this can also be removed,
for example by specifying the prior distribution as uniform with respect to
the Hausdorff measure, as is done in section~\ref{sec:network-example}.

\section{Embedded manifolds with explicit geodesics}
\label{sec:embedded-manifolds}

In this section, we provide examples of embedded manifolds for which the explicit
forms for the geodesic flow is known, and derive the bases for the
normal to the tangent space.

\subsection{Affine subspaces}
\label{sec:affine}

If the embedded manifold is flat, \eg an affine subspace of
$\R^n$, then the geodesic flows are the straight lines
\begin{equation*}
  [x(t), v(t)] = [x(0), v(0)]
  \begin{bmatrix}
    1 & 0 \\
    t & 1
  \end{bmatrix}
  .
\end{equation*}
In the case of the Euclidean manifold $\R^n$, then the normal space to the
tangent is null and no projections are required, and hence the algorithm
reduces to the standard leapfrog scheme of HMC.

In standard HMC, it is common to utilise a ``mass'' or ``preconditioning''
positive-definite matrix $M$, in order to reduce the correlation between
samples, especially where variables are highly correlated or have different
scales of variation. This is directly equivalent to using the RMHMC algorithm with
constant a Riemannian metric, or our geodesic procedure on the embedding of $x
= L^\top q$, where $L$ is a matrix square-root such that $LL^\top = M$ (such
as the Cholesky factor).

\subsection{Spheres}
\label{sec:spheres}

Recall from earlier examples, the unit $(d-1)$-sphere $\sphere^{d-1}$ is an $(d-1)$-dimensional manifold embedded in
$\R^d$, characterised by the constraint
\begin{equation*}
  x^\top x = 1,
\end{equation*}
with tangent space 
\begin{equation*}
  \{v \in \R^d : x^\top v = 0\}.
\end{equation*}
Distributions on spheres, particularly $\sphere^1$ and $\sphere^2$, arise
in many problems in directional statistics \parencite{mardia2000}: examples
include the von Mises--Fisher distribution (Example~\ref{exm:von-mises}), and
the Bingham--von Mises--Fisher distribution (section~\ref{sec:bvmf}). For many
of these distributions, the normalisation constants of the density functions
are often computationally intensive to evaluate, which makes Monte Carlo
methods particularly attractive.

As mentioned in Example~\ref{exm:sphere-geodesic}, the geodesics of the sphere
are the great circle rotations about the origin. The geodesic flows are then
\begin{equation}
  \label{eq:sphere-geodesic}
  [x(t), v(t)] = [x(0), v(0)] 
  \begin{bmatrix}
    1 & 0 \\ 0 & \alpha^{-1}
  \end{bmatrix}
  \begin{bmatrix}
    \cos (\alpha t) & - \sin(\alpha t) \\
    \sin(\alpha t) & \cos (\alpha t)
  \end{bmatrix}
  \begin{bmatrix}
    1 & 0 \\ 0 & \alpha
  \end{bmatrix}
\end{equation}
where $\alpha = \|v(t)\|$ is the (constant) angular velocity. The normal to
the tangent space at $x$ is $x$ itself, so $(I-x x^\top)u$ is an orthogonal
projection of an arbitrary $u \in \R^d$ onto the tangent space.

Other than the evaluation of the log-density and its gradient, the
computations only involve vector-vector operations of addition and
multiplication, so the algorithm scales linearly in $d$.

\subsection{Stiefel manifolds}
\label{sec:stiefel-manifolds}

A \emph{Stiefel manifold} $\stiefel_{d,p}$ is the set of $d\times p$ matrices $X$ such that
\begin{equation*}
  X^\top X = I.
\end{equation*}
In other words, the set of matrices with orthonormal column vectors, or equivalently,
the set of $p$-tuples of orthogonal points in $\sphere^{d-1}$. It is a
$[dp-\tfrac{1}{2}p(p+1)]$-dimensional manifold, embedded in $\R^{d \times
  p}$. 
In the special case where $d = p$, the Stiefel manifold is the \emph{orthogonal
group} $\mathbb{O}_d$: the set of $d \times d$ orthogonal matrices.

These arise in the statistical problems related to dimension reduction such as
factor analysis and principal component analysis, where the aim is to find a
low dimensional subspace that represents the data. They can also arise in
contexts where the aim is to identify orientations, such as projections in shape
analysis, or the eigendecomposition of covariance matrices.

Previously suggested methods of sampling from distributions on Stiefel
manifolds, such as \textcite{hoff2009} and \textcite{dobigeon2010}, have
relied on columnwise Gibbs updates. Such an approach is limited to cases where
the conditional distribution of the column has a conjugate form, and requires
the computation of an orthonormal basis for the null space of $X$,
requiring $O(d^3)$ operations.

Again, we can find the constraints on the phase space by the time-derivative
of the constraint for an arbitrary curve $X(t)$ in $\stiefel_{d,p}$
\begin{equation*}
  \D{t} [X(t)^\top X(t)] = \dot X(t)^\top X(t) + X(t)^\top \dot X(t) = 0.
\end{equation*}
That is, the tangent space at $X$ is the set 
\begin{equation*}
  \{V \in \R^{d \times p} : V^\top X + X^\top V = 0 \} .
\end{equation*}

If we let $\tilde x$ denote the matrix $X$ written as a vector in $\R^{dp}$ by
stacking the columns $x_1,\ldots, x_p$, then an orthonormal basis $N$ for the
normal to the tangent space has $p$ vectors of the form
\begin{equation*}
  \begin{bmatrix}
    x_1 \\ 0 \\ \vdots \\ 0
  \end{bmatrix},
  \begin{bmatrix}
    0 \\ x_2 \\ \vdots \\ 0
  \end{bmatrix},
  \ldots ,
  \begin{bmatrix}
    0 \\ 0 \\ \vdots \\ x_p
  \end{bmatrix}
\end{equation*}
and $\binom{p}{2}$ vectors of the form
\begin{equation*}
  \begin{bmatrix}
    \frac{1}{\sqrt{2}} x_2 \\ \frac{1}{\sqrt{2}} x_1 \\ 0 \\ \vdots \\ 0
  \end{bmatrix}, 
  \begin{bmatrix}
    \frac{1}{\sqrt{2}} x_3 \\ 0 \\ \frac{1}{\sqrt{2}} x_1 \\ \vdots \\ 0
  \end{bmatrix},
  \begin{bmatrix}
    0 \\ \frac{1}{\sqrt{2}} x_3 \\ \frac{1}{\sqrt{2}} x_2 \\ \vdots \\ 0
  \end{bmatrix}
  \ldots
\end{equation*}

For an arbitrary vector $\tilde u \in \R^{dp}$, the projection on to the
tangent space is then
\begin{equation*}
  \tilde u - N N^\top \tilde u =
  \begin{bmatrix}
    u_1 - x_1 (x_1^\top u_1) - \frac{1}{2} x_2 (x_1^\top u_2 + x_2^\top u_1) -
    \ldots \\
    u_2 - x_2 (x_2^\top u_2) - \frac{1}{2} x_1 (x_2^\top u_1 + x_1^\top u_2) -
    \ldots \\
    \vdots
  \end{bmatrix}
\end{equation*}
This can be more easily written in matrix form: for an arbitrary $U \in \R^{d
  \times p}$, the orthogonal projection onto the Stiefel manifold is
\begin{equation*}
  U - \frac{1}{2} X (X^\top U + U^\top X) .
\end{equation*}

The geodesic flows are more complicated than the spherical
case. For $p>1$, they are no longer simple rotations, but can be expressed in
terms of matrix exponentials \parencite[page 310]{edelman1999}
\begin{equation*}
  [X(t), V(t)] = [X(0), V(0)] 
  \exp\left\{ t 
    \begin{bmatrix}
      A & -S(0) \\ I & A
    \end{bmatrix}
  \right\}
  \begin{bmatrix}
    \exp\{-t A\} & 0 \\
    0 & \exp\{-t A\}
  \end{bmatrix}
  ,
\end{equation*}
where $A = X(t)^\top V(t)$ is a skew-symmetric matrix that is constant over
the geodesic, and $S(t) = V(t)^\top V(t)$ is non-negative definite.

Although matrix exponentials can be quite computationally expensive, we note
that the largest exponential of these is of a $2p \times 2p$ matrix, which
requires $O(p^3)$ operations. Other than this and the evaluations of the
log-density and its gradients, all the other operations are simple matrix
additions and multiplications, the largest of which can be done in $O(dp^2)$
operations, hence the algorithm scales linearly with $d$.

For the orthogonal group $\orthonormal_d$ the geodesics have the simpler
form \parencite[equation 2.14]{edelman1999}
\begin{equation*}
  [X(t), V(t)] = [X(0), V(0)] 
  \begin{bmatrix}
    \exp\{t A\} & 0 \\
    0 & \exp\{t A\} 
  \end{bmatrix}
  .
\end{equation*}
As $A$ is skew-symmetric, the Rodrigues' formula gives an explicit form of
$\exp\{t A\}$ when $d=3$ in terms of simple trigonometric functions, and this
can be extended into higher dimensions \parencites{gallier2002}{cardoso2010}.

\subsection{Product manifolds}
\label{sec:product-manifolds}

Given two manifolds $\mathcal{M}_1$ and $\mathcal{M}_2$, their cartesian
product 
\begin{equation*}
  \mathcal{M}_1 \times \mathcal{M}_2 
  = \bigl\{ (x_1,x_2) : x_1 \in \mathcal{M}_1,x_2\in \mathcal{M}_2 \bigr\}
\end{equation*}
is also a manifold. 

Product manifolds arise naturally in many statistical
problems, for example extensions of the von Mises distributions to
$\sphere^1 \times \sphere^1$ (a torus) have been used to model molecular
angles \parencite{singh2002}, and the network eigenmodel in section
\ref{sec:network-example} has a posterior distribution on $\stiefel_{m,p}
\times \R^p \times \R$.

The geodesics of a product manifold are of the form $(\gamma_1,\gamma_2)$,
where each $\gamma_i$ is a geodesic of $\mathcal{M}_i$. Likewise, the tangent
vectors are of the form $(v_1,v_2)$, where each $v_i$ is a tangent to
$\mathcal{M}_i$. Consequently, for an arbitrary vector $(u_1,u_2)$, the
orthogonal projection onto the tangent space is $([I-N_1 N_1^\top] u_1, [I-N_2
N_2^\top] u_2)$, where $N_i$ is an orthonormal basis of $\mathcal{M}_i$.

As a result, when implementing our geodesic Monte Carlo scheme on a product
manifold, the key operations (addition of gradient, projection, and geodesic
update) can be essentially be done in parallel, the only operations requiring
knowledge of the other variables being the computation of the log-density and
its gradient. Moreover, when tuning the algorithm, it is possible to choose
different $\epsilon$ values for each constituent manifold, which can be
helpful when variables have different scales of variation.

\section{Illustrative examples}
\label{sec:examples}

\subsection{Bingham--von Mises--Fisher distribution}
\label{sec:bvmf}

The Bingham--von Mises--Fisher (BVMF) distribution is the exponential family
on $\sphere^{d-1}$ with linear and quadratic terms, with density of
the form
\begin{equation*}
  \pi_\haus (x) \propto \exp\{ c^\top x + x^\top A x \},
\end{equation*}
where $c$ is a vector of length $d$, and $A$ is a $d \times d$ symmetric
matrix \parencite[section 9.3.3]{mardia2000}. 

The Bingham distribution arises as the special case where $c=0$: this is an
axially-bimodal distribution, with the modes corresponding to the eigenvector
of the largest eigenvalue. The BVMF distribution may or may not be bimodal,
depending on the parameter values.

\textcite{hoff2009} develops a Gibbs-style method for sampling from BVMF
distribution by first transforming $y = E^\top x$, where $E^\top \Lambda E$ is the
eigen-decomposition of $A$. Each element $y_i$ of $y$ is updated in random
order, conditional $u \in \sphere^{d-2}$, where $u_j = y_j / \sqrt{1-y_i^2}$
for $j \ne i$. The $y_i \mid u$ are sampled using a rejection sampling scheme with a
beta envelope, however as noted by \textcite{brubaker2012}, this can give
exponentially poor acceptance probabilities (of the order of $10^{-100}$)
 for certain parameter values, particularly when $c$ is large in the direction of
 the negative eigenspectra.

Implementing our geodesic sampling scheme for the BVMF distribution is straightforward,
as the gradient of the log-density is simply $c + 2Ax$, and extremely fast to
run, with run times that are independent of the parameter values. However, as
with any gradient-based method, it has difficulty switching between 
multiple modes; see Figure~\ref{fig:bvmf-single}. 

\begin{figure}
  \centering
  \begin{tikzpicture}
    \begin{groupplot}[group style={group size=3 by 2, 
        y descriptions at=edge left,horizontal sep=0.5cm}, 
      height=4cm,width=5cm, no markers,
      ymin=-1,ymax=1,ylabel={$x_5$},y label style={rotate=-90},
      xmin=0,xmax=200,xticklabel=\empty]
      \nextgroupplot[xlabel={$c_1=0$}] 
      \addplot file {bvmf_0.dat}; 
      \nextgroupplot[xlabel={$c_1=40$}] 
      \addplot file {bvmf_40.dat}; 
      \nextgroupplot[xlabel={$c_1=80$}] 
      \addplot file {bvmf_80.dat};
    \end{groupplot}
  \end{tikzpicture}
  \caption{Trace plots of $x_5$ from 200 samples from the spherical geodesic
    Monte Carlo sampler (with parameters $\epsilon=0.01,T=20$) for a
    Bingham--von Mises--Fisher distribution, with parameters $A =
    \diag(-20,-10,0,10,20)$ and $c=(c_1,0,0,0,0)$. When the distribution is
    bimodal ($c_1 = 0,40$) the sampler has difficulty moving between the
    modes.}
  \label{fig:bvmf-single}
\end{figure}
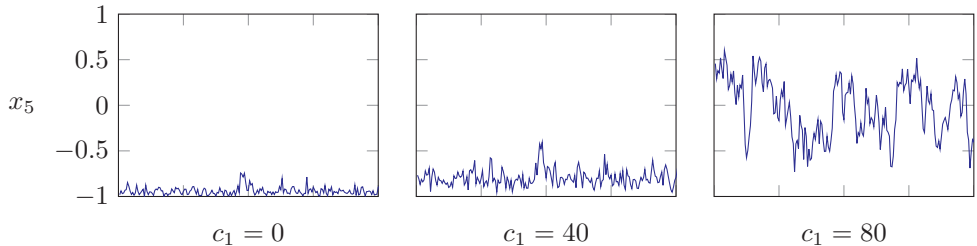
\begin{figure}
  \centering
  \begin{tikzpicture}
    \begin{groupplot}[group style={group size=3 by 2, 
        y descriptions at=edge left,horizontal sep=0.5cm}, 
      height=4cm,width=5cm, no markers,
      ymin=-1,ymax=1,ylabel={$x_5$},y label style={rotate=-90},
      xmin=0,xmax=200,xticklabel=\empty]
      \nextgroupplot[xlabel={$c_1=0$}]
      \addplot file {bvmf_pt_0.dat};
      \nextgroupplot[xlabel={$c_1=40$}]
      \addplot file {bvmf_pt_40.dat}; 
      \nextgroupplot[xlabel={$c_1=80$}]
      \addplot file {bvmf_pt_80.dat};
    \end{groupplot}
  \end{tikzpicture}
  \caption{Trace plots of a simulated tempering scheme applied to the target
    of Figure~\ref{fig:bvmf-single}, using 10 parallel chains to transition
    between multiple modes. The values of
    $\rho$ were $(0.1,0.2,\ldots,1.0)$ and 10 random exchanges were applied
    between parallel geodesic Monte Carlo updates.}
  \label{fig:bvmf-pt}
\end{figure}
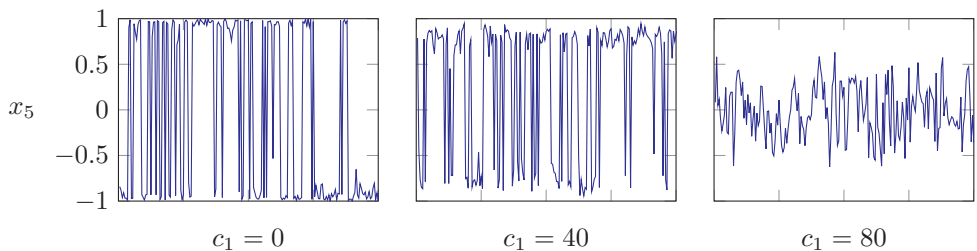

A common method of alleviating this problem is to utilise tempering
schemes \parencite[section 5.5.7]{neal2011}: these operate by sampling from a
class of ``higher temperature'' distributions with densities of the form
\begin{equation*}
  [\pi_\haus(x)]^\rho \quad \text{where $0\leq \rho \leq 1$}.
\end{equation*}
Note that this constitutes a simple linear scaling of the log-density, and so
can be easily incorporated into our method. \emph{Parallel
  tempering} \parencites{geyer1991}[section 10.4]{liu2008} utilises multiple
chains, each targeting a density with a different temperature. The scheme
operates by alternately updating the individual chains, which can be performed
in parallel, and randomly switching the values of neighbouring chains with a
Metropolis--Hastings correction to maintain detailed balance. The results of
utilising such a scheme are shown in Figure \ref{fig:bvmf-pt}

\subsection{Non-conjugate simplex models}
\label{sec:part-observed}

We can use the transformation to the sphere to sample from distributions on
the simplex $\Delta^{d-1}$. These arise in many contexts, particularly as
prior and posterior distributions for discrete-valued random variables such as
the multinomial distribution. 

If each observation $x$ from the multinomial is completely observed, then the
contribution to the likelihood is then $\theta_x$, giving a full likelihood of
at most $d$ terms of form
\begin{equation*}
  L(\theta) = \prod_{i=1}^d \theta_i^{N_i}
\end{equation*}
which is conjugate to a Dirichlet prior distribution.

Complications arise if observations are only partially observed. For example
we may have \emph{marginal} observations, which are only observed to 
a set $S$, in which case the likelihood term is $\sum_{s \in S} \theta_s$, or
\emph{conditional} observations, where the sampling was constrained to
occur within a set $T$, with likelihood term $ \theta_x / (\sum_{t \in T}
\theta_t)$. These terms destroy the conjugacy, and make computation very difficult.

Such models arise under a \emph{case-cohort design} \parencite{lepolain2012},
the risk factors of a particular disease: for the case sample, the risk
factors are observed conditional on the person having the
disease, and for the cohort sample the risk factors are observed marginally (as
disease status is unknown), and overall population statistics may provide some
further information as to the marginal probability of the disease.

The \texttt{hyperdirichlet} R package \parencite{hankin2010} provides an
interface and examples for dealing with this type of data. We consider the
volleyball data from this package: the data arises from a sports league for 9
players, where each match consists of two disjoint teams of players, one of
which is the winner. The probability of a team $T_1$ beating $T_2$ is assumed
to be
\begin{equation*}
  \frac{\sum_{t \in T_1} p_t}{\sum_{t \in T_1 \cup T_2} p_t}
\end{equation*}
where $p=(p_1,\ldots,p_9) \in \Delta^8$. We compare three different methods in
sampling from the posterior distribution for $p$ under a Dirichlet$(\alpha\mathbf{1})$
prior, for different values of $\alpha$. The results are presented in
Table~\ref{tab:simplex}.
\begin{table}
  \centering
  \small
  \tabcolsep=0.1ex
  \begin{tabular}{@{}lSScSScSScSS@{}}
    \toprule
    & \multicolumn{2}{c}{$\alpha = 0.1$} & \
    & \multicolumn{2}{c}{$\alpha = 0.5$} & \ 
    & \multicolumn{2}{c}{$\alpha = 1.0$} & \ 
    & \multicolumn{2}{c}{$\alpha = 5.0$} 
    \\
    \cmidrule{2-3} \cmidrule{5-6} 
    \cmidrule{8-9} \cmidrule{11-12}
    & {ESS \%} & {ESS/s} & &   {ESS \%} & {ESS/s} &
    & {ESS \%} & {ESS/s} & &   {ESS \%} & {ESS/s} 
    \\
    \midrule
    RW-MH 
    & 0.0064 & 6.10 & & 0.113 & 71.1 & 
    & 0.36 & 158 & & 0.84 & 290 \\
    Spherical RW 
    & 0.0089 & 2.48 & & 0.143 & 37.6 & 
    & 0.19 & 51 & & 0.45 & 123 \\
    Simplex HMC 
    & 0.0034 & 0.0079 & & 0.037 & 0.12 & 
    & 53.4 & 611 & & 75.6 & 976 \\
    Spherical HMC 
    & 0.0187 & 0.327 & & 77.3 & 1374 &
    & 92.6 & 1616 & & 187.4 & 3262 \\
    \bottomrule
  \end{tabular}
  \caption{Average effective sample size (ESS) across coordinates per 100
    samples, and per second, of the Volleyball model under a
    Dirichlet$(\alpha\mathbf{1})$ prior from 1\,000\,000 samples. For all
    samplers, $\epsilon = 0.01$, and for the HMC algorithms, $T=20$ integration
    steps were used. We attempted some tuning of the parameters, but were unable to
    obtain any noticeable changes in performance.}
  \label{tab:simplex}
\end{table}

The first is a simple random-walk Metropolis--Hastings algorithm. To ensure
the planar constraint $\sum_{i=1}^9 p_i = 1$ is satisfied, the proposals are
made from a degenerate $\dnorm(x,\epsilon^2 [I-n n^\top])$, where $n =
d^{-1/2} \mathbf{1}$ is the normal to the simplex.

The second is a random walk on the sphere, based on the square root
transformation to the sphere from
Example~\ref{exm:simplex-sphere-embedding}, using proposals of the form
\begin{equation*}
  x_{\text{proposed}} = x \cos(\| \delta\|) + \frac{\delta}{\|\delta\|} \sin(\| \delta\|),
  \quad \text{where} \ 
  \delta \sim \dnorm(0, \epsilon^2[I - x x^\top]).
\end{equation*}
Although this only defines the distribution on the positive orthant, we can
extend this distribution to the entire sphere by reflecting about the axes
(since we only require knowledge of the density up to proportionality, we can
ignore the fact that it is now $2^d$ times larger).  One benefit of this
transformation is that the surface is now smooth and without boundaries, so
proposals outside the positive orthant can be accepted.

The third is the geodesic Monte Carlo algorithm on the simplex. We can ensure the
planar constraint is satisfied via the affine constraint methods in
\ref{sec:affine}, however we need to further ensure that the integration paths
satisfies the positivity constraints, which can be achieved by reflecting the
path whenever it violates the constraint. See the appendix for further details.

The fourth is our proposed geodesic scheme based on the spherical
transformation. As the integrator doesn't pass any boundaries, no reflections
are required.

For small values of $\alpha$, both geodesic Monte Carlo samplers perform
poorly, due to the concentration of the density at the boundaries. These peaks
cause particular problems for the Hamiltonian-type algorithms, as the
discontinuous gradients mean that the integration paths give poor approximations
to the true Hamiltonian paths, resulting in poor acceptance
probabilities. Moreover, for the algorithm on the simplex, the frequent
reflections add to the computational cost.

However when $\alpha=0.5$, the spherical geodesic sampler improves markedly:
recall from Example~\ref{exm:simplex-sphere-jacobian} and
Figure~\ref{fig:beta-circle} that the Dirichlet$(0.5)$ prior is uniform on the
sphere, giving continuous gradients. On the simplex however, the density
remains peaked at the boundaries. The simplex sampler improves considerably
for values of $\alpha \geq 1$ (where the gradient is now flat or negative),
however the spherical algorithm still retains a slight edge. Interestingly, the
spherical random walk sampler performs poorly in all of the examples.

\subsection{Eigenmodel for network data}
\label{sec:network-example}

We use the network eigenmodel of \textcite{hoff2009} to demonstrate how
Stiefel manifold models can be used for dimension reduction, and how our geodesic sampling scheme
 may be used for Stiefel and product manifolds. 
This is a model for a graph on a set of $m$ nodes, where for each unordered
pair of nodes $\{i,j\}$, there is a binary observation $Y_{\{i,j\}}$ indicating
the existence of an edge between $i$ and $j$. 

The specific example of \textcite{hoff2009} is a protein interaction network,
where for $m=270$ proteins, the existence of the edge indicates whether or not
the pair of proteins interact.

The model represents the network by assuming a low ($p=3$) dimensional
representation for the probability of an edge
\begin{equation*}
  P(Y_{\{i,j\}} = 1) = \Phi([U \Lambda U^\top]_{ij} + c),
\end{equation*}
where $\Phi:\R \to (0,1)$ is the probit link function, $U$ is an orthonormal $m
\times p$ matrix, and $\Lambda$ is a $p\times p$
diagonal matrix. $U$ is assumed to have a uniform prior distribution on
$\stiefel_{m,p}$ (with respect to the Hausdorff measure), the diagonal
elements of $\Lambda$ have a $\dnorm(0,m)$ distribution, and $c \sim \dnorm(0,10^2)$.

\textcite{hoff2009} uses columnwise Gibbs updates for sampling $U$, exploiting
the fact that the probit link provides an augmentation that allows these to be
sampled as a Bingham--von Mises--Fisher distribution. However as mentioned in
section~\ref{sec:stiefel-manifolds}, this requires computing the full
null-space of $U$ at each iteration. 
\begin{figure}
  \centering
  \pgfplotstableread{network_a.dat}\networka
  \pgfplotstableread{network_b.dat}\networkb
  \pgfplotstableread{network_pt.dat}\networkpt
  \begin{tikzpicture}
    \begin{axis}[height=6cm,width=12.5cm, no markers,
      ymin=-150,ymax=200,ylabel={$\Lambda_{rr}$},y label style={rotate=-90},
      xmin=0,xmax=1000,xlabel=\empty]
      \addplot[color=lightgreen] table[header=false,x index=0,y index=1]
      {\networkpt};
      \addplot[color=lightgreen] table[header=false,x index=0,y index=2]
      {\networkpt};
      \addplot[color=lightgreen] table[header=false,x index=0,y index=3]
      {\networkpt};
      \addplot[color=navyblue] table[header=false,x index=0,y index=1] {\networka};
      \addplot[color=navyblue] table[header=false,x index=0,y index=2] {\networka};
      \addplot[color=navyblue] table[header=false,x index=0,y index=3] {\networka};
      \addplot[color=midred] table[header=false,x index=0,y index=1] {\networkb};
      \addplot[color=midred] table[header=false,x index=0,y index=2] {\networkb};
      \addplot[color=midred] table[header=false,x index=0,y index=3] {\networkb};
    \end{axis}
  \end{tikzpicture}
  \caption{Trace plots of 1000 samples of the diagonal elements of $\Lambda$
    from geodesic Monte Carlo sampler on the network eigenmodel.  One chain
    (\linesample{navyblue}) converges to the same mode as \textcite{hoff2009},
    while the other (\linesample{midred}) converges to a local mode, with
    approximately $10^{-36}$ of the density. By incorporating this into a
    parallel tempering scheme (\linesample{lightgreen}), the sampler rapidly
    finds the higher mode, and is able to switch between the various permutations.}
  \label{fig:network-example-trace}
\end{figure}
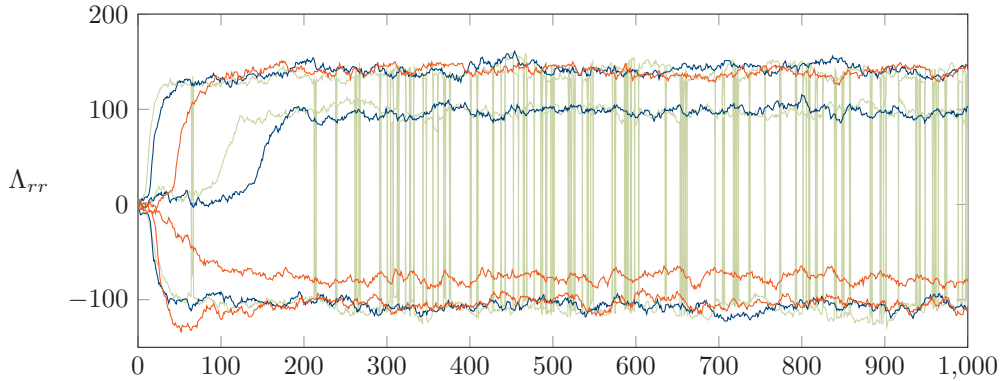

We implement geodesic Monte Carlo on the product manifold, details of which
are given in the appendix. Trace plots from two chains of the diagonal
elements of $\Lambda$ appear in Figure~\ref{fig:network-example-trace}: note
that one chain appears to get stuck in a local mode, while the other converges
to the same as the method of \textcite{hoff2009}. The

By incorporating this approach into a parallel tempering scheme, the model is
able to find the larger mode with greater reliability. Moreover, unlike the
algorithm of \textcite{hoff2009}, it is capable of switching between the
permutations of $\Lambda$, which would further suggest that this is indeed the
global mode.

\section{Conclusion and discussion}
\label{sec:discussion}

We have presented a scheme for sampling from distributions defined on
manifolds embedded in Euclidean space by exploiting their known geodesic
structure.  This method has been illustrated using applications from
directional statistics, discrete data analysis and network analysis. This
method does not require any conjugacy, allowing greater flexibility in the
choice of models: for instance, it would be straightforward to change the
probit link in section~\ref{sec:network-example} to a logit. Moreover, when
used in conjunction with a tempering scheme, it is capable of efficiently
exploring complicated multi-modal distributions.

Our approach could be widely applicable to problems in directional statistics,
such as the estimation of normalisation constants which are often otherwise
numerically intractable. The method of transforming the simplex to the sphere
could be useful for applications dealing with high-dimensional discrete data,
such as statistical genetics and language modelling. Stiefel manifolds arise
naturally in dimension reduction problems, our methods could be particularly
useful where the data are not normally distributed, for instance the analysis
of survey data with discrete responses, such as Likert scale
data. Furthermore, this method could be utilised in statistical shape and
image analysis for determining the orientation of objects in projected images.

The major constraint of this technique is the requirement of an explicit form
for the geodesic flows that can be easily evaluated numerically. These are not
often available, for instance the geodesic paths of ellipsoids require
often computationally intensive elliptic integrals. 

Of the examples we consider, the geodesics of the Stiefel manifold case are
the most demanding, due to the matrix exponential terms. An alternative
approach would be to utilise a Metropolis-within-Gibbs style scheme over
subsets of columns, for example by updating a pair of columns such that they
remain orthogonal to the remaining columns.

However once the geodesics and the orthogonal tangent projection of the
manifold are known, the remaining process of computing the derivatives is
straightforward, and could be easily implemented using automatic
differentiation tools, as is used in the \texttt{Stan} MCMC
library \parencite{stan2012}, currently under development.


\section*{Acknowledgements}
Simon Byrne is funded by a BBSRC grant (BB/G006997) and an EPSRC Postdoctoral
Fellowship (EP/K005723). Mark Girolami is funded by an EPSRC Established
Career Fellowship (EP/J016934) and a Royal Society Wolfson Research Merit
Award.

\subsection*{Contact address}
\noindent \textbf{Email:} \nolinkurl{simon.byrne@ucl.ac.uk}

\noindent \textbf{Postal address:} Department of Statistical Science, University College London,\\
Gower Street, London WC1E 6BT, United Kingdom.

\appendix

\section{Reflecting at boundaries of the simplex}

\textcite[section 5.5.1.5]{neal2011} notes that when an integration path
crosses a boundary of the sample space, it can be reflected about the normal
to the boundary to keep it within the desired space. He considers boundaries
that are orthogonal to the $i$th axis, with normals of the form
$\delta_i$. Whenever such a constraint is violated, the position and velocity
are replaced by
\begin{equation*}
  x'_i = b_i + (b_i - x_i) 
  \quand
  v'_i = - v_i,
\end{equation*}
where $b_i$ is the boundary (either upper or lower). As no other coordinates
are involved in this reflection, this can be done in parallel for all
constrained coordinates.

However for the simplex $\Delta^{d-1}$, the normals are not of the form
$\delta_i$, as this would result in the path being reflected off the plane
$\{x : \sum_i x_i = 1\}$. Instead, we need to reflect about the projection of
$\delta_i$ onto the plane, that is 
\begin{equation*}
  \tilde n_i = \frac{\tilde m_i}{\|\tilde m_i\|} 
  = \frac{d \delta_i - \mathbf{1}}{\sqrt{d(d-1)}}
  \quad \text{where} \quad
  \tilde m_i = \delta_i - (d^{-1/2} \mathbf{1}) (d^{-1/2} \mathbf{1})^\top
  \delta_i .
\end{equation*}

A procedure for performing the position updates is given in
Algorithm~\ref{alg:reflect}.
\begin{algorithm}
\begin{algorithmic}[1]
    \STATE $\omega \leftarrow \epsilon$
    \WHILE{$\omega > 0$}
      \STATE $ (\kappa, j) \leftarrow (\min , \argmin_i) \{ - x_i / v_i : v_i <
        0\} $ 
      \COMMENT{The time until any coordinate is negative-valued: this can only
        occur when the velocity is negative.}
      \STATE $x \leftarrow x + \min(\omega,\kappa) v$
      \STATE $\omega \leftarrow \omega - \min(\omega,\kappa)$
      \IF{$\omega > 0$}
        \STATE $v \leftarrow v - 2 \tilde n_j \tilde n_j^\top v$
      \ENDIF
    \ENDWHILE
\end{algorithmic}
\caption{The geodesic updates on the simplex incorporating reflection off the boundaries.}
\label{alg:reflect}
\end{algorithm}

Unfortunately, this procedure cannot be applied to the RMHMC integrator
proposed by \textcite{girolami2011}, as the implicit steps involved make it
difficult to calculate the reflections.

\section{Network eigenmodel}

Define the $p\times p$ symmetric matrices $\eta = U \Lambda U^\top + c$ and $Y^*$, where
\begin{equation*}
Y^*_{ij} =
\begin{cases}
  1 &\quad Y_{\{i,j\}} = 1 \\
  0 &\quad i = j \\
  -1 &\quad Y_{\{i,j\}} = 0
\end{cases}
.
\end{equation*}
then using the property that $1-\Phi(x) = \Phi(-x)$, the log-density of the
posterior is
\begin{equation*}
  \log \pi_\haus(U,\Lambda,c) = \sum_{\{i,j\}} \log\Phi\bigl( Y^*_{ij} \eta_{ij} \bigr) -
  \sum_{r=1}^p \frac{\Lambda_{rr}^2}{2m} - \frac{c^2}{200} + \text{constant}.
\end{equation*}
The gradients with respect to the parameters are
\begin{align*}
  \Dp[\log \pi_\haus]{U_{ir}} 
  &= \sum_{j=1}^m \Dp[\log \pi_\haus]{\eta_{ij}} U_{jr} \Lambda_{rr}, \\
  \quad
  \Dp[\log \pi_\haus]{\Lambda_{rr}}
  &= 
  \sum_{\{i,j\}} \Dp[\log \pi_\haus]{\eta_{ij}} U_{ir} U_{jr} -
  \frac{\Lambda_{rr}}{m}, \\
  \quad
  \Dp[\log \pi_\haus]{c}
  &= 
  \sum_{\{i,j\}} \Dp[\log \pi_\haus]{\eta_{ij}} -
  \frac{c}{100},
\end{align*}
where the gradients with respect to the linear predictors are
\begin{equation*}
  \Dp[\log \pi_\haus]{\eta_{ij}} 
  = Y^*_{ij} \frac{\phi( Y^*_{ij} \eta_{ij})}{\Phi( Y^*_{ij} \eta_{ij})}.
\end{equation*}

The program was implemented in MATLAB. To avoid numerical overflow errors, the
ratio $\phi(x)/\Phi(x)$, as well as $\log \Phi(x)$ for negative values of $x$,
are calculated using the \texttt{erfcx} function. The matrix exponential terms
were calculated using the inbuilt \texttt{expm} function, which utilises a
Pad\'e approximation with scaling and squaring.

Different $\epsilon$ values were used for each parameter: $\epsilon_U =
0.005$, $\epsilon_\Lambda = 0.1$ and $\epsilon_c = 0.001$. $T=20$ integration
steps were run for each iteration. The parallel tempered version utilised 20
parallel chains, with 10 proposed exchanges between parallel updates.

\printbibliography
\end{document}